# Recent advances in modeling and simulation of biological phenomena in crowded and cellular environments

Apoorva Mathur[1,4], Vanessa Regina Miranda[2,4], Ariane Nunes-Alves[1,3,5]

[1]Institute of Chemistry, Technische Universität Berlin, 10623 Berlin, Germany
[2]Departamento de Físico-Química, Instituto de Química, Universidade Estadual de Campinas, Campinas, São Paulo 13083-861, Brazil
[3]Present address: Brazilian Biosciences National Laboratory, Brazilian Center for Research in Energy and Materials, Campinas, 13083-100, Brazil
[4]These authors are equal first authors.
[5]Corresponding author: ferreira.nunes.alves@tu-berlin.de, ariane.alves@lnbio.cnpem.br

## Abstract
While experiments and computer simulations to study biological phenomena are usually performed in diluted in vitro conditions, such phenomena happen inside the cell, an environment densely packed with diverse macromolecules. Here, we revise recent computational methods to investigate crowded and cellular environments. Protein crowders, inert crowders and small molecules were used to mimic crowding. Simulations were performed for models of the cytoplasm. New methods were developed to simulate crowded systems, reaching up to 200 μs of simulation time. Apart from the challenges, modeling and simulations to investigate biological phenomena inside cells is a growing field, and has a lot of potential to improve our understanding of how such phenomena happen in vivo.

## Introduction

While experiments and computer simulations to study biological phenomena are usually performed in diluted in vitro conditions, such phenomena happen inside the cellular cytoplasm, a more complex and heterogeneous environment. This environment is densely packed with diverse macromolecules, such as proteins and nucleic acids, which can reach concentrations up to 300 g/L in the bacterial cytoplasm [1]. This densely packed environment leads to the phenomenon known as macromolecular crowding, and it can affect biological phenomena via volume exclusion, due to the high concentration of macromolecules or crowders, and via weak and transient interactions of target molecules with crowders, known as soft or quinary interactions [2].

In recent years, several computational methods have been developed and applied to model and simulate macromolecular crowding and cellular environments to further elucidate the effects and consequences of crowding. Such methods build on the pioneering work of McGuffee and Elcock, who modeled and simulated the E. coli cytoplasm [3], and were propelled by the recent surge in methods for integrative biology [4], the increased computational power provided by GPUs and the creation of AlphaFold2 [5] and related tools to model proteins and macromolecular complexes. In the next section, we compare different approaches to model macromolecular crowding. They vary from simple, homogeneous crowding, with a single type of crowder, up to heterogeneous environments mimicking the cytoplasm. In the remaining sections, we present new methods to model and simulate crowding and cellular environments, and summarize the latest mechanistic insights about crowding effects obtained so far.

In this brief article, it is not possible to describe in detail all the relevant methods and systems used to investigate macromolecular crowding. We focused our discussion on publications containing models and simulations with atomistic detail published in 2024 or later, and refer the reader to recent reviews about computational methods to model and simulate crowding and cellular environments [2,6,7]. In comparison to our recent review [2], the present review has a focus on crowded environments and the cytoplasm, and presents significant advances in terms of methods developed and cell types simulated.

## Model systems: from homogeneous environments to bacterial cells and more complex cell types

The cytoplasm is a heterogeneous environment, with a high concentration of macromolecules, and different types of proteins, nucleic acids and metabolites. As an approximation, homogeneous environments, with a high concentration of one type of crowder, have been modeled and simulated (Table 1), resulting in simpler and smaller systems, and allowing one to achieve in-depth mechanistic insights.

Three types of crowders were recently used to mimic macromolecular crowding: (1) protein crowders, such as bovine serum albumin (BSA), lysozyme and ubiquitin [8–13], (2) inert crowders, such as the polymer polyethylene glycol (PEG) [14–17], and (3) small molecules, such as urea and sucrose [8] (Table 1). The choice of such crowders is motivated by the experimental data available for environments crowded with such molecules.

Previous reviews and experimental work [2,18,19] suggest that protein crowders can provide effects similar to heterogeneous cellular environments, while inert crowders are simpler and do not fully capture such effects, as they mimic excluded volume, but not soft interactions with the target protein. However, a recent work [14] challenged the idea that PEG is an inert crowder. Using the fluorescent protein mCherry as a target protein, molecular dynamics (MD) simulations and fluorescence-based experiments, the authors suggest that PEG has soft interactions with the target protein. Another work [8] employed MD simulations to compare the interactions between the target protein SH3 and the protein crowders GB1, BSA, lysozyme and ovalbumin, or the small-molecule crowders urea and sucrose, concluding that urea and sucrose have a different pattern of interaction compared to protein crowders, as they coat the surface of the target protein. More computational work comparing different types of crowders can show whether they have similar effects in comparison to heterogeneous environments.

Models and simulations aimed at describing heterogeneous cellular environments have been explored as an alternative to approaches based on homogeneous systems, in the pursuit of a more comprehensive representation of the cellular cytoplasm (Table 2). The number of such investigations is significantly smaller than that of simulations performed in homogeneous environments, reflecting the computational challenges associated with heterogeneous systems. While in the past it was common to simulate bacterial cells, such as E. coli [3] and Mycoplasma genitalium [20], the area has advanced, and studies now span different types of cells, including E. coli [10] , yeast [21] and human cells (U-2 OS bone osteosarcoma cell, Table 2) [22,23]. There is also large variation in the system sizes, which range from 10 up to 352 proteins in total. Notably, no simulations were performed using coarse-grained representations.

Table 1. Overview of methods, crowders, target protein, total number of crowders in the system, length of simulation, simulation timestep, force field, software, properties observed in simulations of homogeneous environments employing inert crowders, small molecules or protein crowders.

| Method | Crowder | Target proteins | Total number of crowders in the system | Length of simulation (μs) | Simulation timestep (ps) | Force field[a] | Software | Properties observed | Reference |
|---|---|---|---|---|---|---|---|---|---|
| AA-MD[b] | GB1[c]; BSA[d]; OVA[e]; LYZ[f]; urea; sucrose | SH3[g] | GB1: 5-30; BSA: 2-4; OVA: 2-6; LYZ: 3-13; urea: 502; sucrose: 528 | 2 | 0.002 | CHARMM c36m | OpenMM, NAMD | Diffusion | [8] |
| AA-MD | PEG[h] 1,250 | mCherry | 62 | 2 | NA[i] | CHARMM36m | GROMACS | Crowder-protein soft interactions | [14] |
| AA-MD | BSA | Lactate dehydrogenase | 2-8 | 0.52 | 0.0035 (fast motions), 0.007 (slow motions) | AMBER ff99SB-ILDN | GENESIS | Phase separation | [12] |
| CG-MD[j] | PEG 400; PEG 3,000; PEG 8,000 | IDPs[k]: A1-LCD[l]; α-synuclein; Ddx4[m] | PEG 400, PEG 3,000: NA; PEG 8,000: 36- | 20 | 0.01 | CALVADOS 2 | OpenMM | Phase separation | [15] |

| | | | 254 | | | | | | |
|---|---|---|---|---|---|---|---|---|---|
| CG-MD | C60 fullerene; extended rod-like polymer | IDP: Sup35NM[n] | NA | 10 | 20 | Martini 3 | GROMACS | Phase separation | [17] |
| CG-MD | PEG 400;<br>PEG 8,000 | IDPs: ACTR; A1-LCD | 152 | 2.7 | 0.01 | CALVADOS 2 | CALVADOS (OpenMM) | Phase separation | [16] |
| BD[o] | Ribonucleases | PRAI[p] and IGPS[q] | 19 | up to ~18.4 | 0.05 | - | GeomBD3 | Substrate channeling | [13] |
| BD | BSA;<br>LYZ | CI2[r] | BSA: 21-128; LYZ: 96-579 | 5 | 0.5 | - | SDA | Diffusion | [9] |
| Docking and Monte Carlo | BSA;<br>LYZ;<br>CI2 | CI2 | NA | 200 | 20,000 | - | GRAMMCell | Diffusion | [11] |
| Docking and Monte Carlo | System 1: spg[s], UBC[t], VIL1[u]; system 2: ptsH[v], ubiC[w], rnhA[x], map[y], cheY[z]; system 3: spg, UBC, VIL1; system 4: LYZ | System 1: spg, UBC, VIL1; system 2: ptsH, ubiC, rnhA, map, cheY; system 3: spg, UBC, VIL1; system 4: LYZ | System 1: 3,903; system 2: 1,445; system 3: 12,981; system 4: 1,798 | 200 | 20,000 | - | GRAMM | Diffusion | [10] |

**a**-for AA-MD and CG-MD only. **b**-All-atom molecular dynamics. **c**-B1 domain of streptococcal immunoglobulin protein G (with T2Q mutation). **d**-Bovine serum albumin. **e**-Chicken ovalbumin. **f**-Hen egg white lysozyme. **g**-T22G mutant of the N-terminal SH3 domain of drosophila drk. **h**-Polyethylene glycol. **i**-Information not available. **j**-Coarse-grained molecular dynamics. **k**-Intrinsically Disordered Protein. **l**-Low-complexity domain of hnRNPA1. **m**-N-terminal construct of the DEAD-box Helicase 4 protein. **n**-N-terminal and middle regions (NM) of Sup35. **o**-Brownian dynamics. **p**-Phosphoribosyl anthranilate isomerase. **q**-Indoleglycerol phosphate synthase. **r**-chymotrypsin inhibitor 2. **s**-G protein B subunit. **t**-Ubiquitin. **u**-Villin. **v**-ptsH gene. **w**-ubiC gene. **x**-rnhA gene. **y**-map gene. **z**-cheY gene.

Table 2. Overview of methods, cell types, target protein, total number of macromolecules in the system, length of simulation, simulation timestep, force field, software and properties observed in simulations of heterogeneous cellular environments**.**

| **Method** | **Organism / Cell** | **Target proteins** | **Total number of macromolecules in the system** | **Length of simulation (μs)** | **Simulation timestep (ps)** | **Force field**[a] | **Software** | **Properties observed** | **Reference** |
|---|---|---|---|---|---|---|---|---|---|
| AA-MD[b] | Human (U-2 OS bone osteosarcoma cell) | Human PGK[c], yeast PGK | 10-14 | 35 | 0.002 | CHARMM36m | NAMD, Anton 2 software | Protein conformations | [22] |
| AA-MD | Human (U-2 OS bone osteosarcoma cell) | GAPDH[d]; PGK; PGM[e] | 10 | up to 32.6 | 0.002 | CHARMM36m | NAMD, Anton 2 software | Formation of metabolon complexes | [23] |
| BD[f] | Yeast | tRNA[g]; ternary complex (aa-tRNA-EF-1α-GTP)[h] | 138, 204, 205 | 22 - 22.5 | 0.5 | - | SDA | Diffusion | [21] |
| Docking and Monte Carlo | E. coli | - | 352 | 200 | 20,000 | - | GRAMM | Diffusion | [10] |

**a**-for AA-MD only. **b**-All-atom molecular dynamics. **c**-phosphoglycerate kinase. **d**-glyceraldehyde-3-phosphate dehydrogenase. **e**-phosphoglycerate mutase. **f**-Brownian Dynamics. **g**-transfer RNA. **h**-Aminoacyl-transfer RNA – Translational Elongation Factor-1-α - Guanosine Triphosphate.

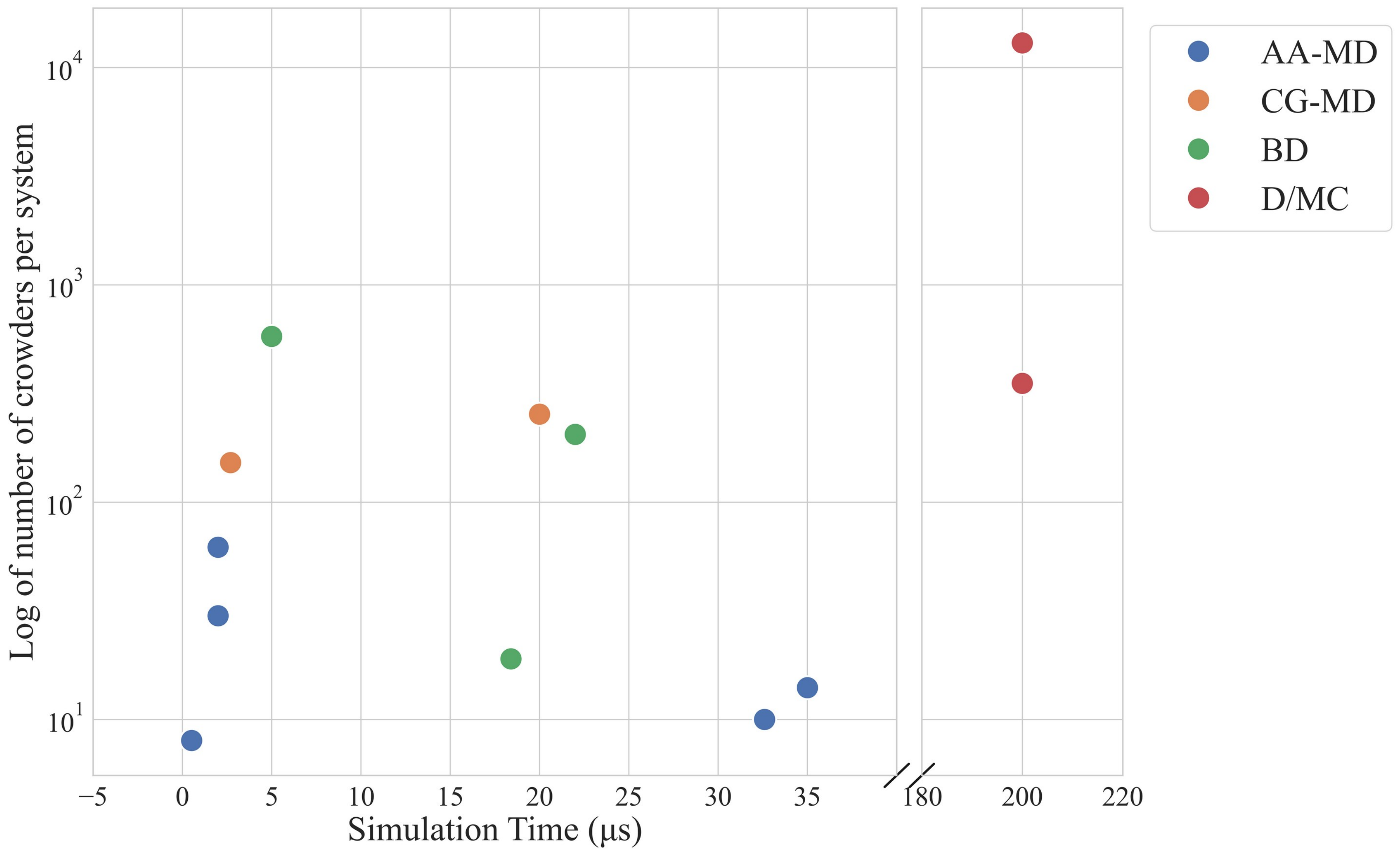

Figure 1. Simulation length depends on the size of the system and molecular-level details of the model. As per the details in Table 1 and Table 2, all-atom molecular dynamics (AA-MD) simulations are performed with smaller systems for longer timescales or with large systems sizes for shorter timescales, while larger systems are simulated with larger timescales with coarse-grained MD (CG-MD), Brownian dynamics (BD) or the protocol combining docking and Monte Carlo (D/MC).

## Novel methods to visualize and simulate crowded and cellular environments

Simulating crowded and cellular environments presents methodological challenges, particularly to balance the system size and representation of spatial heterogeneity with long simulation lengths to reach convergence and independence of initial conditions. Recent methodological advances have focused on alternative sampling strategies and representations, and high-performance computational frameworks that facilitate the simulation of dense environments. Additionally, novel methods were developed to facilitate modeling of cellular environments and analysis of simulations of very large systems, with many particles and macromolecules.

Building and visualizing very large atomistic systems are still quite challenging tasks, and new approaches are under investigation. A new method was developed to create models of bacterial nucleoids, integrating experimental data from different sources, such as cryoelectron tomography, genomic and proteomic data [24]. This method was used to build lattice models and atomic models of the JCVI-Syn3A cell. Another work developed workflows that can convert images and simulations into Minecraft worlds [25], facilitating the visualization of models and simulations of crowded and cellular environments.

From the side of traditional algorithms, MD simulations have been used, in combination with all-atom (AA) or coarse-grained (CG) representations of the systems (Tables 1 and 2). In the CG representation, atoms are grouped in beads, allowing one to simulate larger systems and longer timescales in comparison to the AA representation. However, given the change in free energy barriers and potential energy surfaces in CG, it is not clear if such representation can reproduce experimental time-dependent properties, such as diffusion rates. Additionally, Brownian dynamics (BD) simulations have been employed, where proteins are usually represented as rigid bodies and the solvent is implicit. This allows one to simulate larger systems and to achieve longer timescales, at the expense of more approximations in comparison to AA and CG MD simulations, such as lack of protein flexibility. While protein conformational changes and phase separation are better investigated using AA or CG MD simulations, which include protein flexibility, such methods can be limited in terms of system size and timescales in comparison to BD simulations, which seem to be better suited to simulate the cytoplasm and reproduce associated experimental diffusion rates (Tables 1 and 2). These different approaches can be regarded as complementary, and their use alone or in a multiscale framework depends on the events to be investigated.

From the side of novel algorithms, recent works [26,10,11] developed and applied a protocol combining docking using the GRAMM software [27] with Monte Carlo simulations, which enabled simulations with the largest systems (12981 proteins) and longest timescales (200 μs) among publications from the last two years (Figure 1, Tables 1 and 2). The GRAMMCell web server was developed based on this protocol [11] , facilitating its application. The achievement of such system size and timescale was facilitated by many approximations, such as the treatment of proteins as rigid bodies and the absence of solvent. In order to include time units in the simulation protocol, the authors used diffusion rates from previous MD simulations of villin in a crowded environment. While the protocol, combined with such calibration to include time units, can reproduce previous experimental results, it would be interesting to test if a similar calibration factor would be achieved if other target proteins or experimental results were considered as a reference, and investigate how generalizable this calibration is to other systems. In terms of application, the method is similar to BD simulations, enabling the simulation of larger systems and longer timescales in comparison to AA or CG MD simulations.

Different methodological advances were reported for the GENESIS software, which was developed to simulate biomolecules in solution or in crowded and cellular environments [28–31]. Spatial decomposition analysis (SPANA) is a set of tools developed to facilitate the analysis of trajectories with crowded environments [30], and a new domain decomposition scheme was implemented to improve the efficiency of simulations of crowded environments [28]. Such decomposition scheme can serve as an inspiration to improve parallelization and computational performance in other methods.

Another notable advance is the work from Thornburg et al [32], which integrates different methods, such as BD simulations, a reaction-diffusion master equation, and ordinary differential equations, to simulate the cell cycle in a whole-cell model of the bacterium JCVI-syn3A.

## Effects of crowding on biological phenomena

### A computational microscope to investigate the cytoplasm

Different studies have investigated diffusion, metabolon formation and protein stability inside yeast cells [21] and human cells [22,23]. In BD simulations of the yeast cytoplasm, the heterogeneous environment led to a 8-fold reduction in the diffusion of tRNA and ternary complexes with tRNA, and an increase in macromolecular concentration from 90 g/L to 270 g/L to mimic a cell in osmotic stress further caused a 80-fold reduction in diffusion [21], showing that the excluded volume also modulates diffusion rates. The authors of the same work also performed BD simulations with 70 or 4 different protein types at 90 g/L to mimic the yeast cytoplasm, and observed that tRNA had a slower diffusion rate in the less heterogeneous environment, an effect the authors attributed to the replacement of the ribosome with small proteins, leading to a more even distribution of the excluded volume in the less heterogeneous environment.

In MD simulations of a model of the human cytoplasm, Russell et al observed the formation of a transient metabolon complex involving three enzymes of the glycolytic pathway [23]. Noteworthy, the enzyme-enzyme contacts depended on the force field used. Such simulations provide insight into metabolic pathways functioning and substrate channeling inside cells. It would be interesting to test if the metabolon complex is stabilized or has its formation accelerated by the crowded environment.

### Soft interactions modulate protein diffusion and substrate channeling

The effects of macromolecular crowding on protein diffusion have been examined in several studies [2], mainly due to the availability of experimental diffusion rates to benchmark simulations. There has been growing interest on how variations in crowder properties, such as size, charge, shape, and molecular heterogeneity, affect protein diffusion.

Simulations with the protocol combining docking and Monte Carlo demonstrated that, for the same excluded volume, the diffusion of smaller proteins was more reduced in an environment with self-crowding, in comparison to a heterogeneously crowded environment, while the opposite behavior was observed for larger proteins [10,11]. This result is in agreement with previous experiments [33], and highlights the role of soft interactions and protein size in modulating diffusion rates. Different studies showed that with more crowder-target interactions, diffusion of the target protein decreases, as opposed to in the presence of inert or less interactive crowders [8,9]. Soft interactions with crowders were also reported to slow down substrate channeling between enzymes by up to 8 μs in BD simulations, as the substrate displayed slower diffusion [13]. Such effect is counteracted by the formation of metabolon complexes, which facilitates substrate channeling and may speed up catalysis.

### Crowders can promote or inhibit the formation of condensates and phase separation

Crowders impact the conformation of intrinsically disordered proteins (IDPs), and can be used to promote phase separation and the formation of biomolecular condensates, a topic that received extensive attention in the last years. Here, we focus on simulations of condensates in the presence of crowders, and refer the reader to recent reviews about condensates [34,35]. Rauh et al developed parameters for PEG compatible with the recently developed CALVADOS 2 CG force field, and quantified the phase separation tendencies of three different IDPs, A1-LCD, Ddx4 and α-synuclein, using different PEG sizes and concentrations [15]. While qualitative experimental trends were reproduced, quantitative experimental trends were not fully reproduced for one of the IDPs, indicating that there is room for further refinement of the PEG parameters. Another study investigated the effect of model crowders over phase separation of the prion-like C-terminal domain (TDP-43 CTD) IDP, concluding that while repulsive crowders promote condensation through excluded volume and entropic stabilization, attractive crowders promote condensation through enthalpic interactions [36]. This study modeled crowders as beads, and no comparable experimental results were available to benchmark it. C60 fullerene crowders, which are expected to behave as inert crowders, were used to investigate phase separation of Sup35NM [17] and α-synuclein [37]. All these works used a simplified model of crowder, or a crowder considered inert, highlighting the need to investigate phase separation using more realistic conditions, such as protein crowders. In this direction, a recent work used BSA and PEG to promote phase separation of the enzyme L-lactate dehydrogenase (LDH) [12]. Experiments showed that the catalytic efficiency of LDH was higher in the condensate, while MD simulations suggested that this is due to the stabilization of the closed conformation of the active site of LDH.

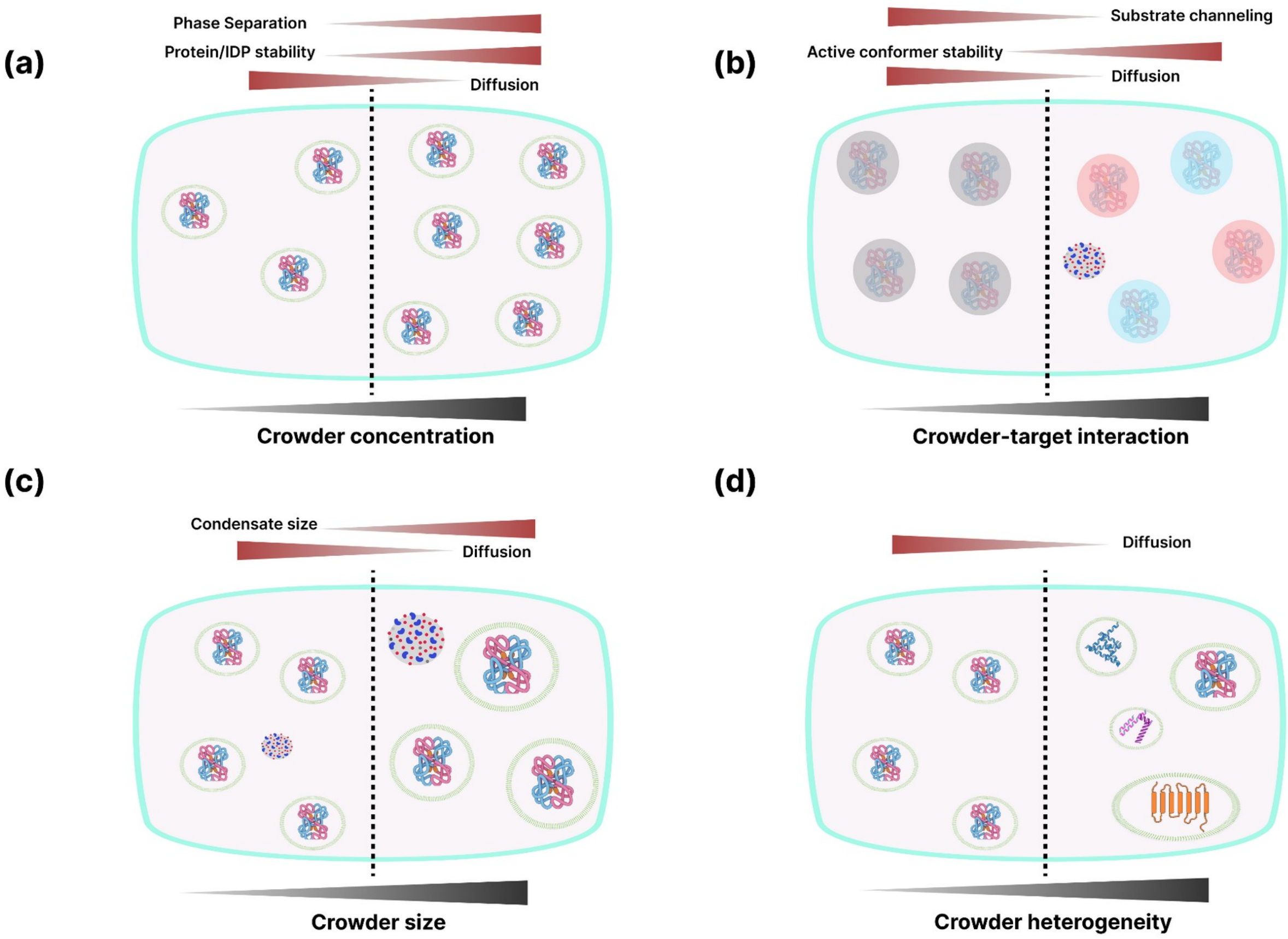

Figure 2. Overview of the effects of macromolecular crowding on proteins and biomolecular condensates. (a) Increasing concentration of crowding agents (pink-blue proteins surrounded by excluded volume in green) generally promotes phase separation, stabilizes proteins, and lowers the diffusion rates of target proteins. (b) Higher interaction between crowder agent and target molecules leads to lower substrate channeling, stabilization of active conformers of proteins and reduction in the diffusion rates of proteins. (c) Increase in crowder sizes promotes coalescence of smaller phase separated droplets into larger ones and reduces the diffusion rate of proteins. (d) Heterogeneous crowder environments were found to be more effective in reducing the diffusion of larger proteins, while smaller proteins had reduced diffusion rates in self crowded homogeneous systems.

## Conclusions and future directions

During the period of the review, different methods were developed and employed to investigate macromolecular crowding, using systems with different degrees of complexity, from single protein or inert crowders to heterogeneous models of the cellular cytoplasm. While many studies investigated protein diffusion, some studies investigated crowding in larger systems, such as the cytoplasm and biomolecular condensates, which shows a trend in simulating environments with growing size and complexity. The investigation of biomolecular condensates, however, usually employed simple inert crowders. Future studies could investigate phase separation in more complex environments, using either protein crowders or portions of the cytoplasm, in combination with further refinement of force fields with experimental data. Due to the diversity of target proteins investigated, in combination with the lack of a clear benchmark system to test new methods, it is challenging to draw broad conclusions about the effects of crowding. The work of Wozniak and Feig [8] partially addresses this issue with the investigation of the effects of different protein crowders on the same target protein, showing that diffusion rates in the presence of crowders are modulated not only by viscosity, but also by the strength of interactions between target protein and crowders.

One of the major challenges in the area is the small amount of experimental data to benchmark simulations, especially for simulations of cellular environments. While previous experimental studies measured diffusion rates of proteins [38] and small molecules [39,40], or probed protein conformational changes inside cells [18,19,41], the area would benefit from more experimental data, in terms of amount and diversity of model systems and techniques employed. One clear need is experimental kinetic rates for protein-protein and protein-ligand binding in crowded environments, which can potentially lead to the development of new strategies for drug design. From the computational side, the area can invest more in the integration of models with structures and data from cryoEM and cryoET experiments, taking the example of tools that better integrate deep learning and cryoET data [42]. As pointed out before [43], another need is the development and adjustment of current simulation software to deal with very large systems containing hundreds of millions of particles, and associated tools for data analysis. The area can also invest in the development of multiscale approaches to build on the strengths of different methods to investigate crowding.

Apart from the challenges, modeling and simulations to investigate biological phenomena inside cells is a growing field, and has a lot of potential to change our understanding of how protein folding, ligand binding and enzyme catalysis happen in vivo, impacting fields such as drug design and biotechnology.

## Declaration of interest

The authors declare no conflict of interest.

## Acknowledgments

The authors thank Dr. Farzin Sohraby (Technical University of Berlin) for suggestions on the manuscript. A.M. and A.N.A. thank funding from DFG under Germany's Excellence Strategy − EXC 2008/1-390540038 − UniSysCat. V.R.M. thanks funding from the BCGE Flexible Travel Funds (Berlin Center for Global Engagement (BCGE) at the Berlin University Alliance (BUA)) and from the Coordenação de Aperfeiçoamento de Pessoal de Nível Superior - Brasil (CAPES) - Finance Code 001. A.N.A. thanks funding from the the Brazilian Biosciences National Laboratory (LNBio), part of the Brazilian Center for Research in Energy and Materials (CNPEM), a private non-profit organization under the supervision of the Brazilian Ministry of Science, Technology, and Innovations (MCTI). The authors acknowledge bioicons for the preparation of Figure 2, under creative commons licenses CC-BY 3.0/4.0: (Protein_quaternary_structure icon and Protein_colored icon by DBCLS (https://creativecommons.org/licenses/by/4.0/), modified 7helix-receptor-membrane icon by Servier (https://creativecommons.org/licenses/by/3.0/); modified biocondensate-concentration icon by Margot Riggi (https://creativecommons.org/licenses/by/4.0/); and cell_membrane_green icon by Helicase 11 (https://creativecommons.org/licenses/by/4.0/)).

**References and recommended reading**

Papers of particular interest, published within the period of review, have been highlighted as:
* of special interest
** of outstanding interest.